\documentclass[english,aps,prb,twocolumn,10pt]{revtex4-1}
\usepackage{lmodern}
\usepackage[T1]{fontenc}
\usepackage[latin9]{inputenc}
\usepackage{babel}
\usepackage{units}
\usepackage{amsmath}
\usepackage{amssymb}
\usepackage{graphicx}
\usepackage{bm}
\usepackage{xcolor}
\usepackage{soul}

\newcommand{\comment}[1]{}

\makeatother

\begin{document}
\title{Spatio-temporal dynamics in graphene}

\author{Roland Jago}
\author{Ra\"ul Perea-Causin}
\author{Samuel Brem}
\author{Ermin Malic}
\affiliation{Chalmers University of Technology, Department of Physics, SE-412 96 Gothenburg, Sweden}
\begin{abstract}
 Temporally and spectrally resolved dynamics of optically excited carriers in graphene has been intensively studied theoretically and experimentally, whereas carrier diffusion in space has attracted much less attention. Understanding the spatio-temporal carrier dynamics is of key importance  for optoelectronic applications, where carrier transport phenomena play an important role. In this work, we provide a microscopic access to the time-, momentum-, and space-resolved dynamics of carriers in graphene. We determine the diffusion coefficient  to be  $D\approx 360 \text{cm}^{2}/s$ and reveal the impact of carrier-phonon and carrier-carrier scattering on the diffusion process. In particular, we show that phonon-induced scattering across the Dirac cone gives rise to back-diffusion counteracting the spatial broadening of the carrier distribution. 
\end{abstract}
\maketitle

The time- and momentum-resolved carrier dynamics in graphene is meanwhile well understood\cite{Malic, butscher07, dawlaty08, plochocka09, Malic2011, Sun2012, AnnalenDerPhysik2017}, but there have been only a few studies on spatio-temporal dynamics and diffusion in graphene \cite{Huang2010,Rengel2014,Ishida2016} and other low dimensional materials, such as carbon nanotubes\cite{Gronqvist2010} and transition metal dichalcogenides\cite{Kato2016,Yuan2017,Rosati2018,Kulig2018}.  Kulig et al. studied\cite{Kulig2018} the exciton diffusion in WS$_{2}$ and determined that the diffusion coefficient varies over two orders of magnitude with respect to the pump fluence. In graphene, pump-probe experiments performed at relatively high pump fluences\cite{Ruzicka2010,Ruzicka2012} demonstrated a diffusion coefficient of  $D=250\pm140\,\text{cm}^{2}/s$ on a picosecond timescale after optical excitation. The diffusion of photoexcited carriers has been studied theoretically\cite{Vasko2012} with an effective Boltzmann approach, where many-particle scattering has been only considered with relaxation rates.\\
However, a full microscopic view on the spatio-temporal dynamics revealing the interplay between diffusion and momentum- and time-dependent scattering processes is still missing. 

\begin{figure}[!t]
\begin{centering}
\includegraphics[width=1\columnwidth]{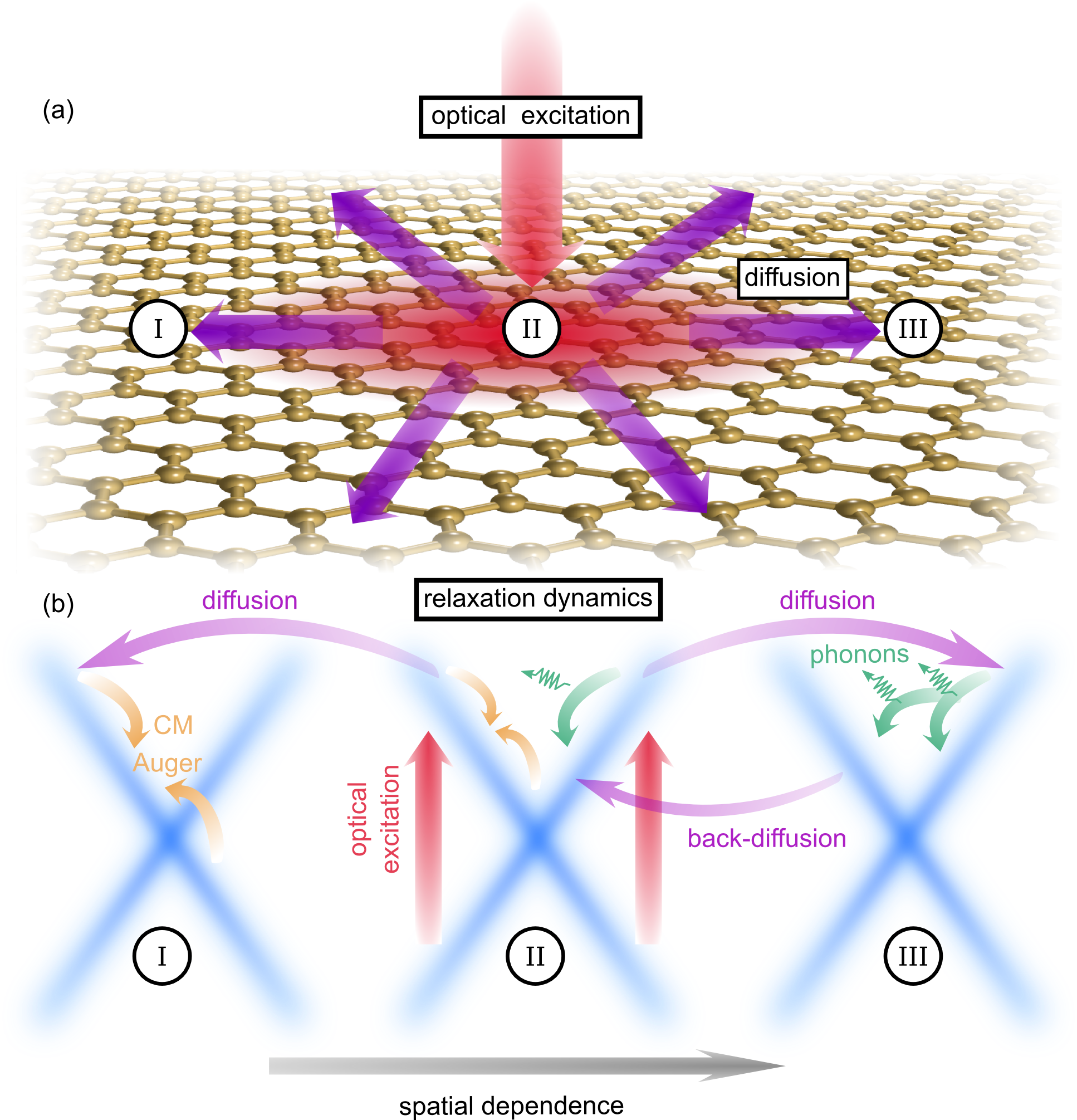} 
\par\end{centering}
\caption{\textbf{Spatio-temporal carrier dynamics in graphene.} (a) Optically excited carriers in spatial region II diffuse to regions I and III. (b) Many-particle scattering leads to relaxation in momentum space in each spatial region. Back-scattering via carrier-phonon processes and the resulting back-diffusion counteracts the spatial distribution of carriers.}
\label{fig:sketch} 
\end{figure}

Exploiting the density matrix formalism \cite{Koch,Kira99} and the Wigner representation\cite{Wigner1932}, we provide microscopic insights into the temporally, spectrally, and spatially resolved dynamics of optically excited carriers in graphene including carrier diffusion, carrier-light, carrier-phonon, and carrier-carrier scattering processes on the same microscopic footing, cf. Fig. \ref{fig:sketch}. In particular, we determine the diffusion coefficient and show that the diffusion process can be tuned with experimentally accessible knobs, such as pump fluence, substrate and temperature. Furthermore, we reveal how carrier-phonon scattering counteracts the diffusion through efficient scattering across the Dirac cone resulting in an efficient back-diffusion, cf. Fig. \ref{fig:sketch}(b). \\

\paragraph*{Theoretical approach:}
We consider a graphene sheet under local optical excitation (red arrows in Fig. \ref{fig:sketch}). The optically excited carriers relax to lower energies via Coulomb- (orange arrows) and phonon-induced scattering (green arrows). The inhomogeneous optical excitation creates spatial gradients in the carrier density, giving rise to diffusion of carriers  (purple arrows).  
To obtain microscopic access to the spatio-temporal dynamics, we derive a set of coupled  equations of motion for the electron occupation probability $\rho_{\mathbf{k}}^{v/c}=\langle a_{\mathbf{k},v/c}^{\dagger}a_{\mathbf{k},v/c}^{\phantom{\dagger}}\rangle$,
the microscopic polarization $p_{\mathbf{k}}=\langle a_{\mathbf{k}v}^{\dagger}a_{\mathbf{k}c}^{\phantom{\dagger}}\rangle$, and the phonon number $n_{\mathbf{q}}^{j}=\langle b_{\mathbf{q}j}^{\dagger}b_{\mathbf{q}j}^{\phantom{\dagger}}\rangle$. Here, the creation and annihilation operators $a_{\mathbf{k}v/c}^{\dagger}$ and $a_{\mathbf{k}v/c}^{\phantom{\dagger}}$ with momentum $\mathbf{k}$ are used for electrons in the valence or conduction band ($v, c$), while the hole occupation probability is given by $\rho_{\mathbf{k}}^{h} = 1 - \rho_{\mathbf{k}}^{v}$.  The corresponding phonon operators are $b_{\mathbf{q}j}^{\dagger}$, $b_{\mathbf{q}j}^{\phantom{\dagger}}$ with the phonon mode $j$ and the phonon momentum $\mathbf{q}$. 

To introduce spatial effects, we transform the occupation probability into the Wigner formalism \cite{Kuhn1996,Rossi2002}. Here, we consider fluctuations of the occupation probability $\rho_{\mathbf{k},\mathbf{q}}^{v/c} = \langle a_{\mathbf{k}-\mathbf{q}/2,v/c}^{\dagger}a_{\mathbf{k}+\mathbf{q}/2,v/c}^{\phantom{\dagger}}\rangle$ and perform the Fourier transformation with respect to the momentum difference $\mathbf{q}$ resulting in the Wigner function
\begin{align}
 f_{\mathbf{k}}^{\lambda}(\mathbf{r}) = \sum_{\mathbf{q}}\,e^{i\mathbf{q}\mathbf{r}}\rho_{\mathbf{k},\mathbf{q}}^{\lambda} 
\end{align}
with $\lambda\!=\!e,h$ denoting electrons in the conduction band and holes in the valence band. Note that the Wigner function $f_{\mathbf{k}}^{\lambda}(\mathbf{r})$ is a quasi-probability function, i.e. $f_{\mathbf{k}}^{\lambda}(\mathbf{r})$ can be negative. Nevertheless, integration over $\mathbf{r}$ or $\mathbf{k}$ gives the actual distribution in momentum  space or the carrier density in real space, i.e. $\rho_{\mathbf{k}}^{\lambda} = 1/A\int\!\! d\mathbf{r}\, f_{\mathbf{k}}^{\lambda}(\mathbf{r})$ or $n({\mathbf{r}}) = 1/A\sum_{\mathbf{k}\lambda}f_{\mathbf{k}}^{\lambda}(\mathbf{r})$ with $A$ as the area of the graphene sheet. 

The carrier dynamics is determined by a many-particle Hamilton operator $H$, where we take into account the free carrier and phonon contribution $H_{\text{0}}$, the carrier-carrier $H_{\text{c-c}}$ and the carrier-phonon $H_{\text{c-ph}}$ interaction accounting for Coulomb- (orange arrows) and phonon-induced scattering (green arrows), and the carrier-light coupling $H_{\text{c-l}}$ (red arrows) that is treated on a semi-classical level. Details on the contributions of the many-particle Hamilton operator including the calculation of the matrix elements can be found in Refs. \onlinecite{Malic,Malic2011}.

Exploiting the Heisenberg equation of motion, we derive the equation of motion for the carrier fluctuation $\rho_{\mathbf{k},\mathbf{q}}^{\lambda}$. Taking into account the free-particle Hamilton operator $H_{0}$ leads to $i\hbar\dot{\rho}_{\mathbf{k},\mathbf{q}}^{\lambda} = \big(\varepsilon_{\mathbf{k}+\mathbf{q}/2}^{\lambda} - \varepsilon_{\mathbf{k}-\mathbf{q}/2}^{\lambda}\big)\rho_{\mathbf{k},\mathbf{q}}^{\lambda}$ with the electronic dispersion $\varepsilon_{\mathbf{k}}^{\lambda}$. To determine an equation for the Wigner function we perform a Fourier transformation resulting in $i\hbar\dot{f}_{\mathbf{k}}^{\lambda}(\mathbf{r}) = \int\!\! d\mathbf{r'}\,\sum_{\mathbf{q}}\big(\varepsilon_{\mathbf{k}+\mathbf{q}/2}^{\lambda} - \varepsilon_{\mathbf{k}-\mathbf{q}/2}^{\lambda}\big)\,e^{i\mathbf{q}\mathbf{r'}}f_{\mathbf{k}}^{\lambda}(\mathbf{r}-\mathbf{r'})$. To simplify this integro-differential equation we expand the Wigner function to the first order $f_{\mathbf{k}}^{\lambda}(\mathbf{r}\!-\!\mathbf{r'}) \approx f_{\mathbf{k}}^{\lambda}(\mathbf{r})-\mathbf{r'}\nabla_{\mathbf{r}}^{\phantom{+}}f_{\mathbf{k}}^{\lambda}(\mathbf{r})$. By using $\mathbf{r'}e^{i\mathbf{q}\mathbf{r'}} = -i\nabla_\mathbf{q}e^{i\mathbf{q}\mathbf{r'}}$ and shifting the $\mathbf{q}$-derivative to the electron dispersion via partial integration, the $\mathbf{r'}$-integral depends only on the exponential function resulting in $\delta_{\mathbf{k},\mathbf{q}}$, whereby the zeroth order of the expansion of the Wigner function vanishes. Finally, the equation of motion for the Wigner function for the free Hamilton operator reads
\begin{align}
 \dot{f}_{\mathbf{k}}^{\lambda}(\mathbf{r},t) = -\frac{1}{\hbar} \nabla_{\mathbf{k}}^{\phantom{+}}\varepsilon_{\mathbf{k}}^{\lambda}  \cdot \nabla_{\mathbf{r}}^{\phantom{+}}f_{\mathbf{k}}^{\lambda}(\mathbf{r},t).
\end{align}
To derive the equations of motion for the Wigner function, the polarization and the phonon number with the full Hamilton operator we make the following assumptions: (i) We consider diffusion processes in the polarization  to be small, since the latter quickly decays in momentum space and vanishes directly after the optical excitation\cite{Malic2011}. In contrast, the relaxation of carriers occurs on a picosecond timescale which is comparable to diffusion processes, and therefore the diffusion term can not be neglected in the equation for the Wigner function. (ii) We also neglect the phonon diffusion, since it is expected to be much slower than the electronic diffusion due to the flat phonon dispersion. (iii) We expect scattering processes between different spatial positions to be small compared to the diffusion. Now, using the Heisenberg equation of motion, we derive the full spatio-temporal graphene Bloch equations in second-order Born-Markov approximation
\begin{align}
 \dot{f}_{\mathbf{k}}^{\lambda}(\mathbf{r}, t) &=\Gamma_{\mathbf{k}\lambda}^{\text{in}}(\mathbf{r}, t)\,\hat{f}_{\mathbf{k}}^{\lambda}(\mathbf{r}, t)-\Gamma_{\mathbf{k}\lambda}^{\text{out}}(\mathbf{r}, t)\,f_{\mathbf{k}}^{\lambda}(\mathbf{r}, t)\label{eq:rho}\\ 
 &\nonumber
 +2\,\text{Im}\big[\Omega_{\mathbf{k}}^{vc, *}(\mathbf{r}, t) p_{\mathbf{k}}^{\phantom{+}}(\mathbf{r}, t)\big]-  \frac{\nabla_{\mathbf{k}}^{\phantom{+}}\varepsilon_{\mathbf{k}}^{\lambda}}{\hbar}\cdot\nabla_{\mathbf{r}}^{\phantom{+}}f_{\mathbf{k}}^{\lambda}(\mathbf{r}, t),\\[5pt]
 \dot{p}_{\mathbf{k}}(\mathbf{r}, t) &= i\Delta\omega_{\bf{k}}(\mathbf{r}, t)p_{\mathbf{k}}(\mathbf{r}, t) \!-\! i\Omega_{\mathbf{k}}^{vc}(\mathbf{r}, t)\bar{f}_{\mathbf{k}}(\mathbf{r}, t),\label{eq:p}\\[5pt]
 \dot{n}_{\mathbf{q}}^{j}(\mathbf{r}, t)&=\Gamma_{\mathbf{q}j}^{\text{em}}(\mathbf{r}, t)\hat{n}_{\mathbf{q}}^{j}(\mathbf{r}, t)\!-\!\Gamma_{\mathbf{q}j}^{\text{ab}}(\mathbf{r}, t)\, n_{\mathbf{q}}^{j}(\mathbf{r}, t)\!-\!\gamma_{\text{ph}}^{\phantom{+}} \bar{n}_{\mathbf{q}}^{j}(\mathbf{r}, t)\label{eq:n}
\end{align}
with the abbreviations $\hat{f}_{\mathbf{k}}^{\lambda}(\mathbf{r}, t)=1-f_{\mathbf{k}}^{\lambda}(\mathbf{r}, t), \bar{f}_{\mathbf{k}}(\mathbf{r}, t)=f_{\mathbf{k}}^{e}(\mathbf{r}, t)\!+\!f_{\mathbf{k}}^{h}(\mathbf{r}, t)\!-\!1, \hat{n}_{\mathbf{q}}^{j}(\mathbf{r}, t)= n_{\mathbf{q}}^{j}(\mathbf{r}, t)\!+\!1$, and $\bar{n}_{\mathbf{q}}^{j}(\mathbf{r}, t)= n_{\mathbf{q}}^{j}(\mathbf{r}, t)-n_{\mathbf{q},\text{B}}^{j}$ with the initial Bose-distribution for phonons $n_{\mathbf{q},\text{B}}^{j}$. 
The equations describe the time-, momentum- and space-resolved coupled dynamics of electrons/holes, phonons, and the microscopic polarization. 
 The dynamics of electrons in the conduction band and holes in the valence band is symmetric, but has different initial conditions for doped graphene samples.
The appearing Rabi frequency is defined as $\Omega_{\mathbf{k}}^{vc}(\mathbf{r},t)=i\frac{e_{0}}{m_{0}}  \mathbf{M}_{\mathbf{k}}^{vc}\cdot \mathbf{A}(\mathbf{r},t)$ with the free electron mass $m_{0}$, the vector potential $\mathbf{A}(\mathbf{r},t)$, and the optical matrix element $\mathbf{M}_{\mathbf{k}}^{vc}=\langle\mathbf{k}v|\nabla_{\mathbf{k}}|\mathbf{k}c\rangle$. Since we study the carrier dynamics close to the Dirac point, renormalization effects can be neglected.
Furthermore, we have introduced $\hbar\Delta\omega_{\mathbf{k}}(\mathbf{r},t)=(\varepsilon_{\mathbf{k}}^{v}-\varepsilon_{\mathbf{k}}^{c} + i\gamma_{\mathbf{k}}(\mathbf{r},t))$ with the electronic dispersion $\varepsilon_{\mathbf{k}}^{\lambda}$ and the dephasing rate $\gamma_{\mathbf{k}}(\mathbf{r},t)$. The time-, momentum- and spatial dependent dephasing  $\gamma_{\mathbf{k}}(\mathbf{r},t)$ and in- and out-scattering rates $\Gamma_{\mathbf{k}\lambda}^{in/out}(\mathbf{r},t)$ include carrier-carrier and carrier-phonon scattering channels. The dynamics of the phonon number $n_{\mathbf{q}}^{j}(t)$ is driven by the emission and absorption rates \cite{Malic,Malic2011} $\Gamma_{\mathbf{q}j}^{em/abs}(\mathbf{r},t)$.
The constant $\gamma_{\text{ph}}$ is the experimentally determined phonon decay rate \cite{Kang2010}. More details on the appearing many-particle scattering and dephasing rates can be found in Refs. \onlinecite{Malic, Malic2011}. 
In this work, we assume that graphene lies on a $\text{SiC}$-substrate  and is surrounded by air on the other side. This is taken into account by introducing an averaged dielectric background constant\cite{Patrick1970} $\varepsilon_{bg}=\frac{1}{2}\big(\varepsilon_{\text{s}}+1\big)$, where $\varepsilon_{\text{s}}= 9.66$ is the static screening constant of the substrate, while 1 describes the dielectric constant of air. Furthermore, the internal many-particle screening is taken into account by calculating the static limit of the Lindhard equation \cite{Kira2006,Koch}, which screens the Coulomb matrix elements. 

The derived set of equations resemble the semiconductor Bloch equations for spatial homogeneous systems (cf. Refs. \onlinecite{Malic,Malic2011}) up to the additional term $\nabla_{\mathbf{k}}^{\phantom{+}}\varepsilon_{\mathbf{k}}^{\lambda}/\hbar \cdot\nabla_{\mathbf{r}}^{\phantom{+}}f_{\mathbf{k}}^{\lambda}(\mathbf{r},t)$, which describes the diffusion of carriers in the direction $\nabla_{\mathbf{k}}\varepsilon_{\mathbf{k}}^{\lambda} \propto \mathbf{e}_{\mathbf{k}} = \mathbf{k}/|\mathbf{k}|$. As  a result, carriers with different sign in momentum move in opposite directions generating locally asymmetric carrier distributions in momentum space and resulting in a local current ${\mathbf{j}}(\mathbf{r},t) = -\frac{4 e_{0}v_{\text{F}}}{A}\sum_{\mathbf{k}\lambda}f^{\lambda}_{\mathbf{k}}(\mathbf{r},t)\,\mathbf{e}_{\mathbf{k}}$ with  the Fermi velocity $v_{\text{F}}$. The sum contains both  electrons in the conduction band and holes in the valence band and in a spatially homogeneous system, the mean current vanishes.

\paragraph*{Spatio-temporal dynamics:}

\begin{figure}[!t]
\centering
\includegraphics[width=1.05\columnwidth]{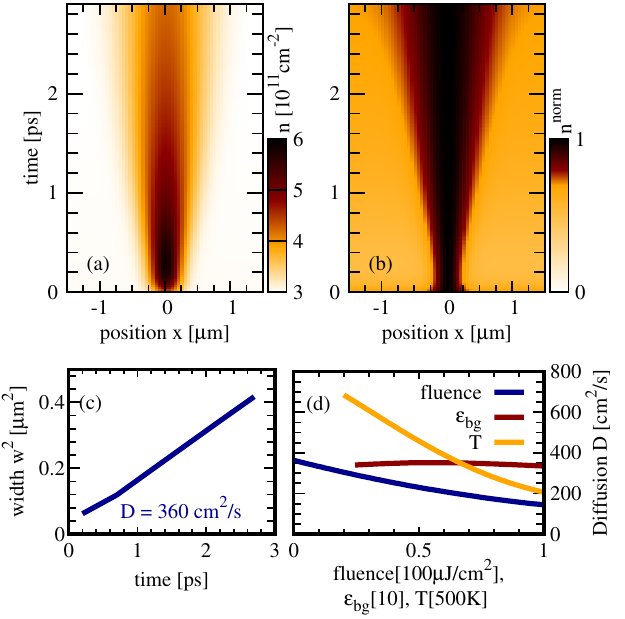} 
\caption{\textbf{Spatio-temporal dynamics.} (a) Carrier density $n(x,t)$ is shown as a function of time and position after a local optical excitation centered at $x=0$.  (b) Carrier density normalized to the maximal density at each time step to highlight the diffusion of carriers. (c) The square of the width $w$ of the spatial distribution is a measure for the diffusion coefficient $D$. (d) Diffusion coefficient as a function of different pump fluences, substrates and temperatures. The $x$-axis is normalized to the maximal value of the respective parameter.}
\label{fig:diffusion} 
\end{figure}

Now, we numerically evaluate the spatio-temporal graphene Bloch equations and investigate the interplay of diffusion and relaxation processes after optical excitation. We excite carriers with an optical pulse with a Gaussian profile both in time and space. We chose typical values for pulse characteristics including a temporal FWHM of $115\,\text{fs}$, a spatial FWHM of $265\,\text{nm}$, an excitation energy of $1\,\text{eV}$ and a pump fluence of $1\mu\text{J/cm}^{2}$. The temporally and spatially dependent carrier density $n(x, t)$ is shown in Fig. \ref{fig:diffusion} (a).  The diffusion of carrriers is reflected in the broadening of the carrier density in space.  Normalizing the density for each time step, the broadening becomes more visible (Fig. \ref{fig:diffusion} (b)), since  phonon- and Auger-driven interband processes give rise to a reduction of carriers with increasing time. 

To quantify the diffusion and to estimate the diffusion coefficient for graphene, we fit  the carrier density with a Gaussian $\text{exp}(-x^2/w^2(t))$ for every time step. The temporal evolution of the width $w(t)$ is depicted in Fig. \ref{fig:diffusion} (c). It is connected to an effective diffusion coefficient $D$ via\cite{Pathria} $w^{2}(t)=w^{2}_{0}+4D\,t$ resulting in $D\approx 360\,\text{cm}^{2}/s$ for the investigated graphene sample on a SiC substrate. Our results fit well to the experimentally obtained values\cite{Ruzicka2012} for the diffusion coefficient of $D=250\pm140\,\text{cm}^{2}/s$. The obtained values for the diffusion coefficient can be also translated into an effective mobility $\mu$ by using the Einstein relation\cite{Einstein1905} $\mu = e_{0}D/(k_{B}T)$. At room temperature, we obtain a carrier mobility of approximately $14000\,\text{cm}^{2}/\text{Vs}$ which is in the range of experimentally reported values\cite{Farmer2011, Zhu2009}.
In Fig. \ref{fig:diffusion} (d) we show the influence of  pump fluence, substrate and temperature on the diffusion coefficient. We find that the temperature has the largest impact. The underlying processes will be discussed below.

\begin{figure}[!t]
\centering
\includegraphics[width=0.8\columnwidth]{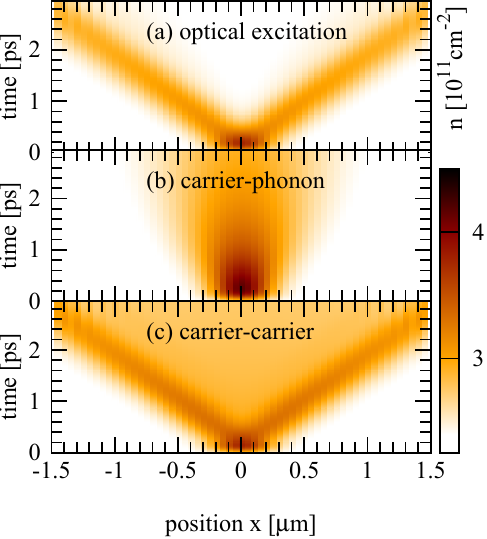} 
\caption{\textbf{Impact of scattering channels on diffusion.} (a) After optical excitation the carriers on Dirac cone branches with different sign of momenta diffuse in opposite directions. (b) Carrier-phonon scattering counteracts the diffusion and the broadening of the distribution due to back-scattering between Dirac cone branches and the following back-diffusion (Fig. \ref{fig:channels} (c)). (c) Carrier-carrier scattering does not effectively counteract the diffusion, since parallel scattering is preferred by the Coulomb matrix elements. Thus, one can still clearly observe the spatial separation of carriers with opposite momentum.}
\label{fig:channels} 
\end{figure}

Now, we investigate the impact of different scattering mechanisms on the diffusion process, cf. Fig. \ref{fig:channels}. We start with the case without any scattering channels just considering the electron-light interaction. After the optical excitation, carriers with positive/negative momenta diffuse in opposite  spatial directions according to the diffusion term in Eq. (\ref{eq:rho}).  After approximately $100\,\text{fs}$ the carrier separation becomes visible, as the intial carrier density distribution splits into two pronounced peaks of the same width but with half of the amplitude, cf. Fig. \ref{fig:channels} (a). Including the carrier-phonon scattering, we observe a strongly reduced spatial broadening of the carrier density and no splitting appears  (Fig.\ref{fig:channels} (a)). Phonon-induced relaxation processes counteract the diffusion via back-scattering across the Dirac cone and the following back-diffusion (cf. Fig. \ref{fig:sketch}). The impact of carrier-phonon scattering will be further microscopically resolved in the next section. Including only the carrier-carrier scattering,  the density diffuses with the same speed as in the case without any scattering channels (cf. Fig. \ref{fig:channels} (c)). This is a consequence of the symmetry of Coulomb matrix elements,  which favor parallel scattering\cite{Malic2012,Mittendorff2014}. Scattering across the Dirac cone is relatively inefficient and back-scattering is even forbidden. In contrast to the case without scattering, the spatial region between the two peaks contains a non-zero density. This  reflects the weak but not vanishing Coulomb scattering processes bringing carriers from one to the other side of the Dirac cone.

\begin{figure}[!t]
\centering
\includegraphics[width=1.0\columnwidth]{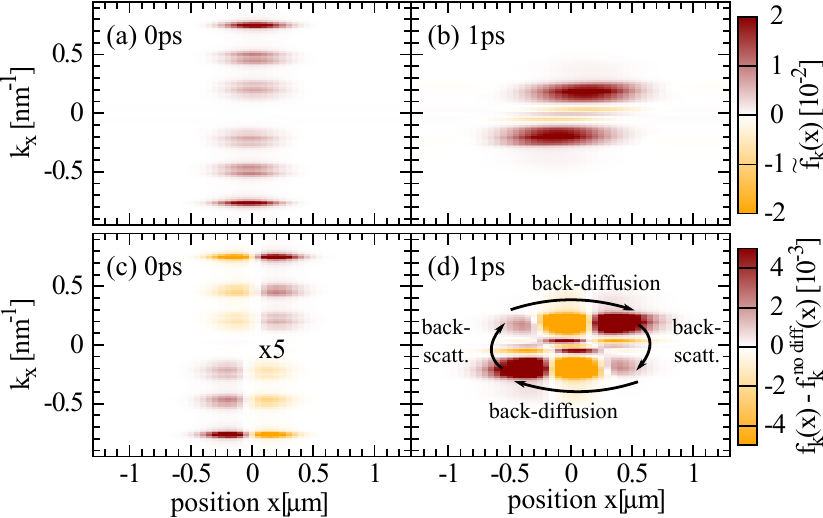} 
\caption{\textbf{Impact of carrier-phonon scattering.} (a)-(b) Wigner function is shown in dependence of space and momentum for two fixed times. Time 0 ps corresponds to the maximum of the optical excitation pulse. Note that we plot here the change in the Winger function with respect to the intial time $t_0$ before the optical excitation, i.e. $\tilde{f}_{k}(x,t) = f_{k}(x,t) - f_{k}(x,t_0)$. (c)-(d) Illustration of the impact of diffusion by showing the Wigner function minus the case without diffusion. The overall spatial carrier distribution becomes broader at larger times due to diffusion. Backscattering with acoustic phonons leads to additional peaks at lower energies. Since here the momentum of the carriers is flipped, they diffuse back resulting in multiple sign changes in (d).}
\label{fig:phonon} 
\end{figure}

\textit{Carrier-phonon dynamics}: To get a thorough understanding of the microscopic processes governing the spatio-temporal carrier dynamics, we investigate the spectral and spatial behaviour of the Wigner function for different times. We start with the interplay of diffusion and carrier-phonon scattering processes. The optically excited carriers scatter via optical phonons to lower energies and form enhanced carrier occupations separated by the energy of optical phonons (red regions in Fig. \ref{fig:phonon} (a)). Diffusion processes lead to a spatial broadening of the carrier distribution and after approx. $1\,\text{ps}$ the carriers have relaxed to lower energies close to the Dirac point (Fig. \ref{fig:phonon} (b)).

To investigate the impact of diffusion in more detail we performed the same calculation twice, but in the second computation we excluded diffusion processes. Illustrating the difference of both calculations, i.e. i.e.  $f_{k}(x) - f_{k}(x)^{\text{no diff}}$, we can directly observe the impact of diffusion on carrier-phonon scattering (Figs. \ref{fig:phonon} (c)-(d)). As already discussed in the theory section carriers with positive/negative momentum diffuse in opposite spatial direction. This behaviour is illustrated in Fig. \ref{fig:phonon} (c), where carriers with positive momentum diffuse from $x<0$ positions (orange spots) to $x>0$ positions (red spots). After $1\,\text{ps}$ the carriers have already relaxed to energies close to the Dirac cone and below the optical phonon energy. Consequently, the scattering with acoustic phonons becomes dominant. Due to the flat dispersion of acoustic phonons with respect to the Dirac cones back-scattering across the Dirac cone is preferred, such that carriers with positive momenta are scattered to negative momenta and vice versa (Fig. \ref{fig:sketch}). The inversion of momenta results in a back-diffusion, such that the overall carrier distribution stays bunched in space, cf. Fig. \ref{fig:channels} (b). The back-diffusion is shown in Fig. \ref{fig:phonon} (d) by the the multiple sign change in the colored regions (red to orange to red).\\

\begin{figure}[t!]
\centering
\includegraphics[width=1.0\columnwidth]{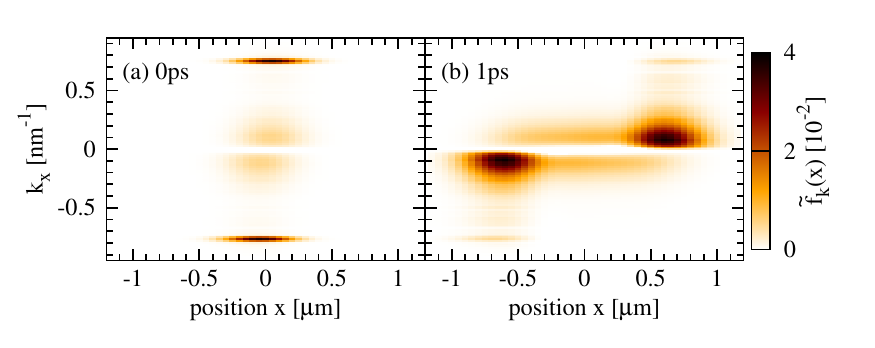} 
\caption{\textbf{Impact of carrier-carrier scattering.} The same as Fig. \ref{fig:phonon}, but now only including the carrier-carrier instead of carrier-phonon scattering. Coulomb interaction is most efficient for parallel scattering along the Dirac cone in the momentum space. As a result, carrier-carrier scattering does not efficiently counteract the diffusion of carriers in opposite direction as clearly observed in (b).}
\label{fig:coulomb} 
\end{figure}

\textit{Carrier-carrier dynamics}: Now, we investigate the impact of carrier-carrier scattering on diffusion of optically excited carriers. An important aspect here is that Auger scattering is efficient giving rise to a carrier multiplication\cite{Winzer2010_Multiplication,Brida2013,Ploetzing2014,05_Mittendorff_Auger_NatPhys_2014,Gierz2015} that increases the overall carrier density (note the scale of the color map in Fig. \ref{fig:coulomb} compared to Fig. \ref{fig:phonon}(a)). This also results in a quick increase of the carrier distribution close to the Dirac cone already during the optical excitation (Fig. \ref{fig:coulomb}(a)). Since electrons and holes diffuse in the same direction, the conditions for carrier multiplication are still satisfied after the diffusion.
The directional dependence (in momentum space) for intraband carrier-carrier scattering is determined by the Coulomb matrix element that includes a form factor proportial to\cite{Malic} $ 1+e^{i\varphi}$ with the scattering angle $\varphi$. This means that parallel scattering ($\varphi=0$) is the preferable scattering channel, and that for the back-scattering ($\varphi=\pi$) the amplitude of the Coulomb matrix element completely vanishes. As a result, scattering processes across the Dirac cone that change the sign of the carrier momentum (and lead to a back-diffusion) are inefficient. As a result,  carriers with positive/negative momenta remain separated with respect to their spatial position - similarly to the case without any scattering (Fig. \ref{fig:channels}(a)).  Figure \ref{fig:coulomb} (b) illustrates that carriers  with positive/negative momenta are mainly distributed towards positive/negative spatial positions.

\textit{Tuning the diffusion:} Now, we can explain the dependence of the diffusion coefficient on pump fluence, substrate and temperature shown in Fig. \ref{fig:diffusion} (d). We find that the diffusion becomes less efficient with the increasing pump fluence (blue curve). Here,  more carriers are excited resulting also in an increased number of emitted phonons. Thus, hot-phonon effects become important, i.e. an increasing number of phonons can be reabsorbed in back-scattering processes giving rise to additional channels for back-diffusion.
Furthermore, we find that the diffusion coefficient is nearly independent of the substrate (red curve) entering in our calculations through the screening of the Coulomb potential. This is not surprising, since Coulomb-induced scattering processes have been shown to only play a minor role for the diffusion of carriers, cf. Fig. \ref{fig:channels}. Finally, we observe that the diffusion can be most efficiently tuned by varying the temperatures (orange curve). The lower the temperature, the weaker the  carrier-phonon scattering, the less efficient is back-scattering and back-diffusion resulting in a considerably increased diffusion coefficient.\\
In summary, we provide a microscopic view on the spatio-temporal carrier dynamics in graphene based on the density matrix formalism in Wigner representation. We investigate the interplay of diffusion and many-particle scattering processes after a local optical excitation. In particular, we determine a diffusion coefficient of $D\approx 360\text{cm}^{2}/s$ that agrees well with recent experimental values. Furthermore, we reveal that carrier-phonon scattering across the Dirac cone and the resulting back-diffusion are crucial ingredients to understand the spatial broadening of the carrier distribution. The gained insights are important e.g. for graphene-based photodetectors \cite{Koppens2014,sun2014,Buscema2015,Mueller2016,Levitov2011}, that are governed by the thermoelectric effect, which relies on spatial temperature gradients.\\
This project has received funding from the European Union's Horizon 2020 research and innovation programme under grant agreement No. 696656 (Graphene Flagship). Furthermore, we acknowledge support from the Swedish Research Council (VR). The computations were performed on resources at Chalmers Centre for Computational Science and Engineering (C3SE) provided by the Swedish National Infrastructure for Computing (SNIC).

\bibliographystyle{apsrev4-1}

\end{document}